\documentclass[
    ,final            
  ]{aipproc}
\usepackage[OT1]{fontenc}
\usepackage[latin1]{inputenc}
\layoutstyle{8x11single}
\usepackage{epsfig} 
\usepackage{graphicx}
\usepackage{psfrag}
\usepackage{amssymb, amsmath, amsfonts}
\usepackage{epstopdf}
\newcommand{\ww}[1]{\underline{\underline{{\bf #1}}}}
\newcommand{\ave}[1]{\langle #1 \rangle}
\newcommand{\sti }{\ww{K}}
\newcommand{\dep}{{\bf U}}                                          

\begin{document}

\title{How granular materials deform in quasistatic conditions}


\author{J.-N. Roux
}{
  address={Universit\'e Paris-Est, Laboratoire Navier, 2 All\'ee Kepler, Cit\'e Descartes,
  77420 Champs-sur-Marne, France}
}
\author{G. Combe
}{
  address={Laboratoire 3SR, Universit\'e Joseph Fourier,
  38041 Grenoble, France}
}

\begin{abstract} 
Based on numerical simulations of quasistatic deformation of model granular materials, 
two rheological regimes are distinguished, according to whether macroscopic strains merely 
reflect microscopic material strains within the grains in their contact regions (type I strains), 
or result from instabilities and contact network rearrangements at the microscopic level (type II strains).
We discuss the occurrence of regimes I and II in simulations of model materials made of disks (2D) or spheres (3D). The transition from regime I to regime II in monotonic tests such as triaxial compression is different from both the elastic limit and from the yield threshold. The distinction between both types of response is shown to be crucial for the sensitivity to contact-level mechanics, the relevant variables and scales to be considered in
micromechanical approaches, the energy balance and the possible occurrence of macroscopic instabilities 
\end{abstract}

\maketitle


\section{Introduction}
\subsection{The quasistatic limit, the rigid limit and the macroscopic limit} 
Although they are modeled, at the macroscopic level, with constitutive laws in which 
physical time and inertia play no part~\cite{GDV84,DMWood}, 
granular materials are most often investigated at the 
microscopic level by ``discrete element'' numerical methods (DEM) in which the motion of the 
solid bodies is determined through integration of dynamical equations involving masses and accelerations.
Fully quasistatic approaches, in which the system evolution in configuration space, as some loading parameter is varied, is regarded as a continuous set of mechanical equilibrium states, are quite rare in the numerical
literature~\cite{RC02,KTKK03,McHe06}. It is regarded as a natural starting point, on the other hand, 
to perform suitable averages of the mechanical response of the elements of a contact network to derive the
macroscopic material response~\cite{LRMJ09}. Whether  and in which cases it is possible to dispense
with dynamical ingredients of the model at the granular level and how the quasistatic limit is approached are
fundamental issues that still need clarification.

Another set of open questions are related to the role of particle deformability. Most DEM studies include contact elasticity in the numerical model. Experimentally, elastic behavior is routinely measured in
quasistatic tests~\cite{Tat104} and sound propagation. Yet, most often, contact deflections are quite negligible in comparison with grain diameters. In the ``contact dynamics'' method~\cite{JEA99,RaRi09}, which is used to simulate quasistatic granular rheology~\cite{RR04,RTR04,R08}, grains are modeled as rigid, undeformable solid bodies. The influence of contact deformability on the macroscopic behavior, the existence of a well-defined rigid limit are thus other basic issues calling for further investigations. 

Small granular samples, as the ones used in DEM studies, often exhibit quite noisy mechanical properties. The approach to a macroscopic behavior expressed with smooth stress-strain curve might seem problematic, especially
in the presence of rearrangement events, associated with instabilities at the microscopic level~\cite{CR2000,staron02b}.

\subsection{The origins of strain}
The present communication shows how one may shed light on the
interplay between the quasistatic, rigid and macroscopic limits on distinguishing
two different rheological regimes and delineating their conditions of occurrence, in simple
model materials.
Macroscopic strain in solidlike granular materials has two obvious
physical origins: first, grains deform near their contacts, where
stresses concentrate (so that one models the grain interaction
with a point force); then, grain packs rearrange as contact networks
break, and then repair in a different stable configuration. We refer here respectively to the two
different kinds of strains as type I and II. The 
present paper, based on numerical simulations of simple materials, identifies the regimes, denoted as I and II
accordingly, within which one mechanism or the other dominates, and discusses
the consequences on the quasistatic rheology of granular materials. 
\section{Numerical model materials and simulation procedures}\label{sec:basic}
Two sets of numerical simulation results are exploited below.
Two-dimensional (2D) assemblies of polydisperse disks, as in Refs.~\cite{Gael2,RC02,CR03}, are subjected
to fully stress-controlled biaxial tests, for which a quasistatic computation method~\cite{Gael2,RC02,RC09} is exploited, in addition to standard DEM simulations. 
The behavior of three-dimensional (3D) packs of monosized spherical particles, as in Refs.~\cite{Roux05,iviso1,iviso3} 
is studied in simulated triaxial compression tests, with special attention to strains in the quasistatic limit. 
Part of the results are presented in the references (mostly in some conference proceedings) cited just above, pending the publication of a more comprehensive study.
\subsection{Two-dimensional material and stress-controlled tests}
\begin{figure}[!htb]
\centering
\includegraphics[width=6.5cm]{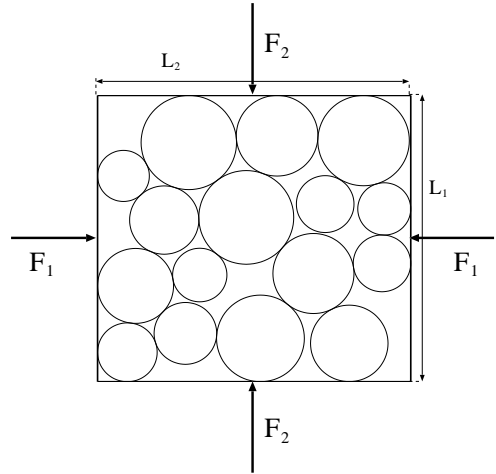}
\caption{ 
Schematic representation of the biaxial tests simulated on 2D disk samples. $\sigma_2=F_2/L_2$ is kept constant, equal to the initial isotropic pressure $P$, while $\sigma_1 = F_1/L_1$ is stepwise increased. 
\label{fig:schemabiaxial}}
\end{figure}

2D systems are simulated in order to investigate basic rheophysical mechanisms with good accuracy, in
the simplest conceivable, yet representative, model material.
Samples made of polydisperse disks in 2D, with a uniform diameter distribution between $0.5a$ and $a$, are first
assembled on isotropically compressing frictionless particles, thus producing dense packs with solid fraction 
$\Phi = 0.8434\pm 3\times 10^{-4}$ and coordination number $z$ close to 4 in the large system limit. Those values are extrapolated from data averaged on sets of 
samples with $N=1024$, $3025$ and 4900 disks. The samples are enclosed in a rectangular cell framed by
solid walls, 2 of which are mobile orthogonally to their direction, which enables us to carry out biaxial compression tests (Fig.~\ref{fig:schemabiaxial}). Finite system effects on $\Phi$ and $z$ are mainly due
to the surrounding walls and can be eliminated (they are proportional to perimeter to area ratio).
\subsubsection{Stress-increment controlled DEM simulations}
Once prepared in mechanical equilibrium under an isotropic pressure $P$, disk samples, in which contacts are now regarded as frictional, with friction coefficient $\mu=0.25$, are subjected to biaxial
tests as sketched in Fig.~\ref{fig:schemabiaxial}.
Strains $\epsilon_1 = -\Delta L_2/L_2$, $\epsilon_2 = -\Delta L_1/L_1$, ``volumetric'' strain  
$\epsilon_v= \epsilon_1+\epsilon_2$ are measured in equilibrium configurations, while the stress deviator $q=\sigma_2-\sigma_1$ is the control parameter. We use soil mechanics conventions, for which compressive
stresses and shrinking strains are positive. $q$ is stepwise increased by small intervals $\Delta q = 10^{-3}P$.
The contact model is the standard (Cundall-Strack~\cite{CS79}) one with normal ($K_N$) and tangential ($K_T$) stiffness constants such that $K_N=2K_T=10^5P$. A normal viscous force is also introduced in the contact, in order to reach equilibrium configurations faster. After each deviator step, one waits for the next equilibrium configuration, in which forces and moments are balanced with good accuracy (with a 
tolerance below $10^{-5} Pa$ for forces on grains, below $10^{-5} PL$ for forces on walls). We refer to this procedure as \emph{stress-increment controlled} (SIC) DEM.
\subsubsection{The stricly quasistatic approach: SEM calculations}
\begin{figure}[!htb]
\centering
\begin{minipage}[c]{.35\linewidth}
{
  \psfrag{k}{$K_N$}
  \psfrag{g}{\hskip -0.35cm$\eta_N$}
  \includegraphics[height=3.5cm]{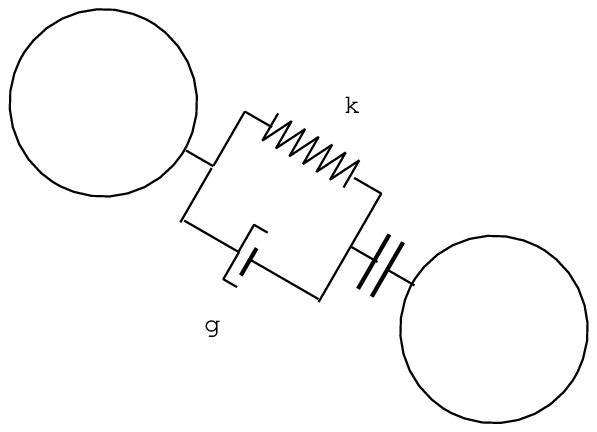}
}
\end{minipage}
\hskip 1.5cm
\begin{minipage}[c]{.5\linewidth}
{
  \psfrag{k}{$K_T$}
  \psfrag{m}{\hskip -0.3cm $\mu$}
  \includegraphics[height=3.5cm]{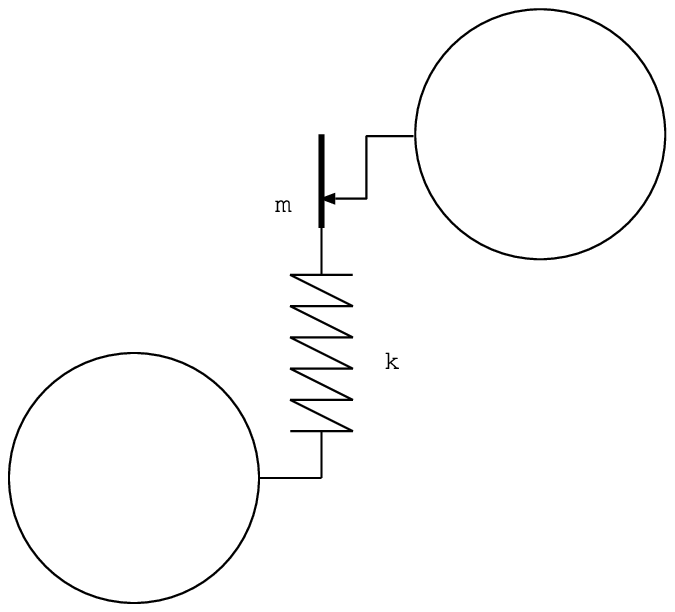}
}
\end{minipage}
\caption{\label{fig:rheonet} Normal (left) and tangential (right) contact behavior in 2D disk samples, as schematized with rheological elements: springs with stiffness constants $K_N$, $K_T$, dashpot with damping constant $\eta_N$, plastic slider with threshold related to normal force by coefficient $\mu$.}
\end{figure} 
The \emph{static elastoplastic method} (hereafter referred to as SEM), amounts to dealing with the
initial sample configuration as a network of springs and plastic sliders corresponding to contact behavior, as in
Fig.~\ref{fig:rheonet} -- with the dashpots ignored, as they play no role in statics. The evolution of
the system under varying load is determined as a continuous trajectory in configuration space, each point of
which is an equilibrium state. It has been implemented in~\cite{Gael2,RC02}, and a similar approach was used in~\cite{KTKK03}. The algorithm will not be described here, as it is presented in~\cite{RC09}. It relies on 
resolution of linear system of equations, with the form of the matrix (the elastoplastic stiffness matrix) depending on contact status (nonsliding, sliding, open). The bases of the approach are also discussed in~\cite{McHe06}.

SEM calculations are possible as long as only type I strains are obtained, and the results reported here~\cite{Gael2} correspond to the deviator interval $0\le q\le q_1$ in biaxial compressions from the chosen
initial state (in which all tangential forces are equal to zero), in which a type I response is obtained.
\subsection{Triaxial compression of 3D bead assemblies} 
Triaxial compression tests of assemblies of $N=4000$ single-sized 
spherical beads of diameter $a$ are simulated by DEM, with the more standard procedure in
which the \emph{axial strain rate} $\dot\epsilon _a$ is kept constant (hereafter \emph{strain-rate controlled} or
SRC DEM). The deviator stress, $q$, is measured, as a 
function of axial strain $\epsilon_a=\epsilon_1$, as $q = \sigma_1-\sigma_3$, where $\sigma_1$ is the major (``axial'') principal stress conjugate to $\epsilon_a$, while the other two (lateral) principal stresses $\sigma_2=\sigma_3$ are kept equal to the initial isotropic pressure $P$. To allow for comparisons with laboratory experiments, the
beads are attributed the elastic properties of glass (Young modulus $E=70$~GPa, Poisson ratio $\nu=0.3$) and friction coefficient $\mu=0.3$. The contact law is a somewhat simplified version of the Hertz-Mindlin ones~\cite{JO85}, as in Ref.~\cite{iviso1}, which might be consulted for more details. It leads to favorable
comparisons of elastic moduli~\cite{iviso3} obtained in simulations and measured in experiments on glass beads.
Preparation of cuboidal samples with periodic boundaries in all three directions (and thus statistically
 homogeneous and devoid of wall effects) under prescribed pressures in the range $10$~kPa~$\le P\le 1$~MPa 
is detailed in Refs.~\cite{iviso1,iviso2}. It is shown~\cite{iviso1} that one may obtain, depending on the
assembling procedure, for densities close to the random close packing limit $\Phi\simeq 0.64$ under low pressure, coordination numbers ranging from $z\simeq 4$ 
(or $z^*\simeq 4.5$ if the ``rattlers'', i.e., grains carrying no force, are excluded from the count) to
$z=6$ (more exactly $z^*=6$, with 1 or 2\% of rattlers) in the limit of $P\to 0$. As in \cite{iviso1}, the 
low-coordination systems ($z^*\simeq 4.5$) are referred to as ``C samples'' in the sequel, while those with
$z^*\simeq 6$ are called ``A samples''. We can
thus assess the influence of initial coordination number on the small strain (pre-peak) behavior
of a dense material.
\subsection{Dimensionless control parameters}
The contact law and the simulated mechanical test lead to the definition of useful dimensionless numbers. 
The \emph{inertia parameter} $I = \dot \epsilon_a\sqrt{m/aP}$ (in 3D) or $I = \dot \epsilon_a\sqrt{m/P}$ (in 2D)
characterizes the importance of inertial effects in strain-rate controlled tests under pressure $P$ ($m$ is the grain mass). 
The parameter $I$ has been used repeatedly to describe the state of granular materials in steady flow, both
in experiments~\cite{Gdr04} and in simulations~\cite{Dacruz05,Hatano07,PR08a}, or the
departure from equilibrium in a slow compression~\cite{iviso2}.
It also plays a central role in the recent formulation of a successful constitutive law for dense granular flows~\cite{JFP06}.

The importance of contact deflections, relative to grain diameter $a$, is expressed by the stiffness number, $\kappa$, 
which is defined as $\kappa = K_N/P$ in 2D models with linear contact elasticity, and as $\kappa = \left(\dfrac{E}{(1-\nu^2)P}\right)^{2/3}$ with 3D beads and Hertzian contacts. In both cases, typical contact deflections $h$ satisfy $h/a\propto \kappa^{-1}$~\cite{iviso1}.

The three limits mentioned in the introduction can be defined as $I\to 0$ (quasistatic limit), 
$\kappa\to\infty$ (rigid limit), $N\to\infty$ (macroscopic limit).

\section{SIMULATION RESULTS}
\subsection{Biaxial tests in 2D}
\subsubsection{Type I response interval, quasistatic approach}
\begin{figure}[!htb]
\centering
\includegraphics[angle=270,width=8.5cm]{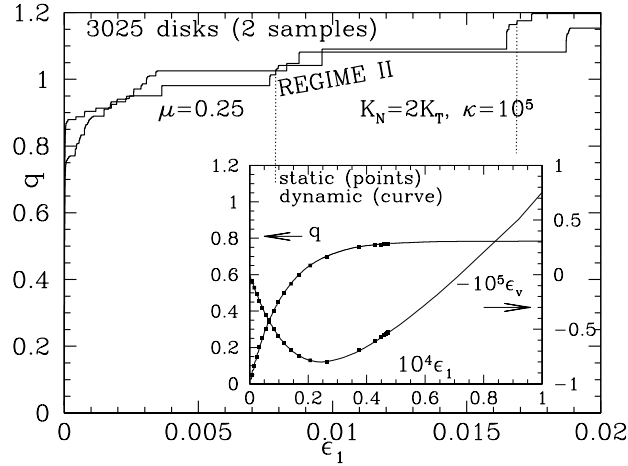}
\caption{$q$ (normalized by $P$) versus $\epsilon_a$ in SIC tests on 2 samples of 3025 disks, showing a very stiff
increase (confused with vertical axis), and then a staircase regime. Inset: detail of very small strains, 
with comparison of SIC DEM and SEM calculations.
\label{fig:dessmarfro}}
\end{figure}
In Fig.~\ref{fig:dessmarfro}, $q(\epsilon_a)$ curves as obtained by SIC DEM are shown for two samples of 3025
disks. The curves first exhibit a very sharp increase of deviator $q$, which, as revealed once the  
strain scale is blown up by a factor of $10^4$ in the insert, is in fact an interval of type I response: direct
SEM calculation are possible, and coincide with DEM results. The smoothness of the stress variations versus strain
in that range is characteristic of a continuous trajectory of equilibrium states in configuration space. 
Beyond the transition to type II strain regimes, a staircase-shaped deviator curve  (Fig.~\ref{fig:dessmarfro}) is observed, 
exhibiting intervals of stability (nearly vertical parts of curve in Fig.~\ref{fig:dessmarfro}), 
separated by rearrangement events (horizontal parts of curve in Fig.~\ref{fig:dessmarfro}) 
in which the system gains kinetic energy before a new stable contact network is formed. We could check that the SEM procedure is 
able to reproduce the stability intervals obtained with SIC DEM. On reversing the load (stepwise decreasing 
$q$), a considerably larger deviator range is accessible to SEM calculations, and thus in regime I,
as illustrated by the two (quasi-vertical) dotted lines on the main
plot of Fig.~\ref{fig:dessmarfro}. 
\subsubsection{Role of contact stiffness}
\begin{figure}[htb!]
 \centering
 \psfrag{y}[r]{$q/p$}
 \psfrag{y2}[l]{$\epsilon_v$}
 \psfrag{2e-05}[l]{\begin{small}$\phantom{0.}2$\end{small}}
 \psfrag{4e-05}[l]{\begin{small}$\phantom{0.}4$\end{small}}
 \psfrag{5e-05}[l]{\begin{small}$\phantom{0.}5$\end{small}}
 \psfrag{3e-05}[l]{\begin{small}$\phantom{0.}3$\end{small}}
 \psfrag{1e-05}[l]{\begin{small}$\phantom{0.}1$\end{small}}
 \psfrag{0}[l]{\begin{small}$0$\end{small}}
 \psfrag{k1}[l]{\begin{small}$ \kappa=10^4$\end{small}}
 \psfrag{k2}[l]{\begin{small}\hskip -9mm$\kappa=10^5$\end{small}}
 \psfrag{x}[l]{\begin{small}$\kappa\epsilon_1$\end{small}}
\includegraphics[width=0.4\textwidth]{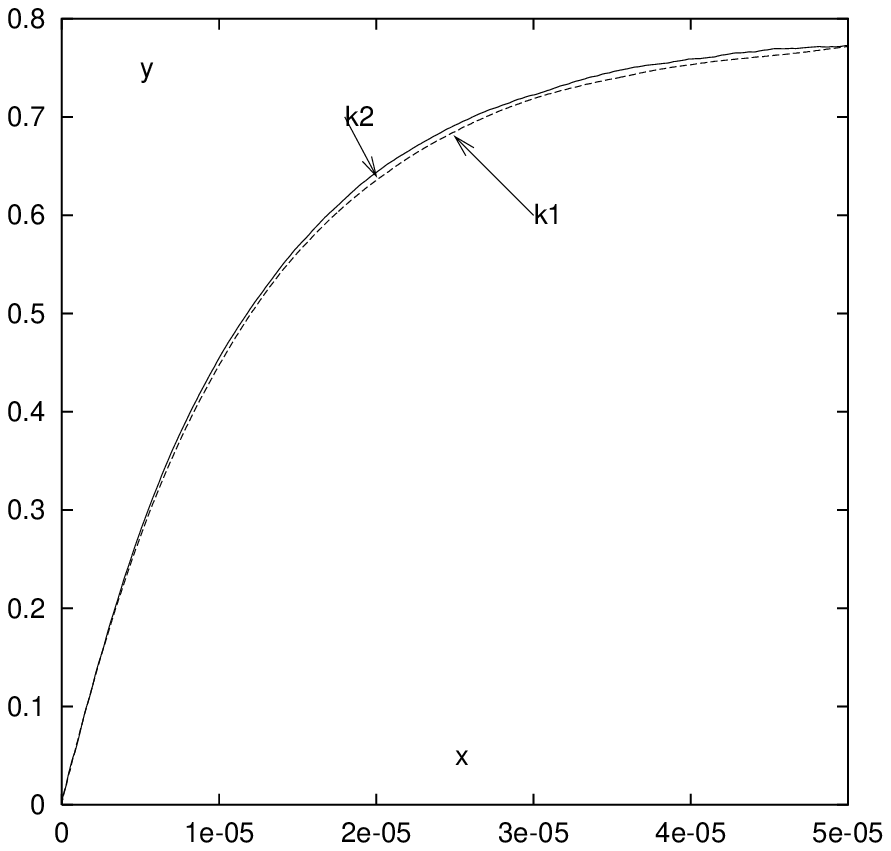}
\hfil
 \psfrag{y}[l]{\begin{small}$\kappa\epsilon_v$\end{small}}
 \psfrag{2e-05}[l]{\begin{small}$\phantom{0.}2$\end{small}}
 \psfrag{4e-05}[l]{\begin{small}$\phantom{0.}4$\end{small}}
 \psfrag{5e-05}[l]{\begin{small}$\phantom{0.}5$\end{small}}
 \psfrag{3e-05}[l]{\begin{small}$\phantom{0.}3$\end{small}}
 \psfrag{1e-05}[l]{\begin{small}$\phantom{0.}1$\end{small}}
 \psfrag{-2e-06}[l]{\begin{small}$-0.2$\end{small}}
 \psfrag{-4e-06}[l]{\begin{small}$-0.4$\end{small}}
 \psfrag{-6e-06}[l]{\begin{small}$-0.6$\end{small}} 
 \psfrag{-8e-06}[l]{\begin{small}$-0.8$\end{small}}
 \psfrag{-1e-05}[l]{\begin{small}$-1$\end{small}}
 \psfrag{0}[l]{\begin{small}$0$\end{small}}
 \psfrag{k1}[l]{\begin{small}$ \kappa=10^4$\end{small}}
 \psfrag{k2}[l]{\begin{small}$\kappa=10^5$\end{small}}
 \psfrag{x}[c]{\begin{small}$\kappa\epsilon_1$\end{small}}
 \includegraphics[width=0.4\textwidth]{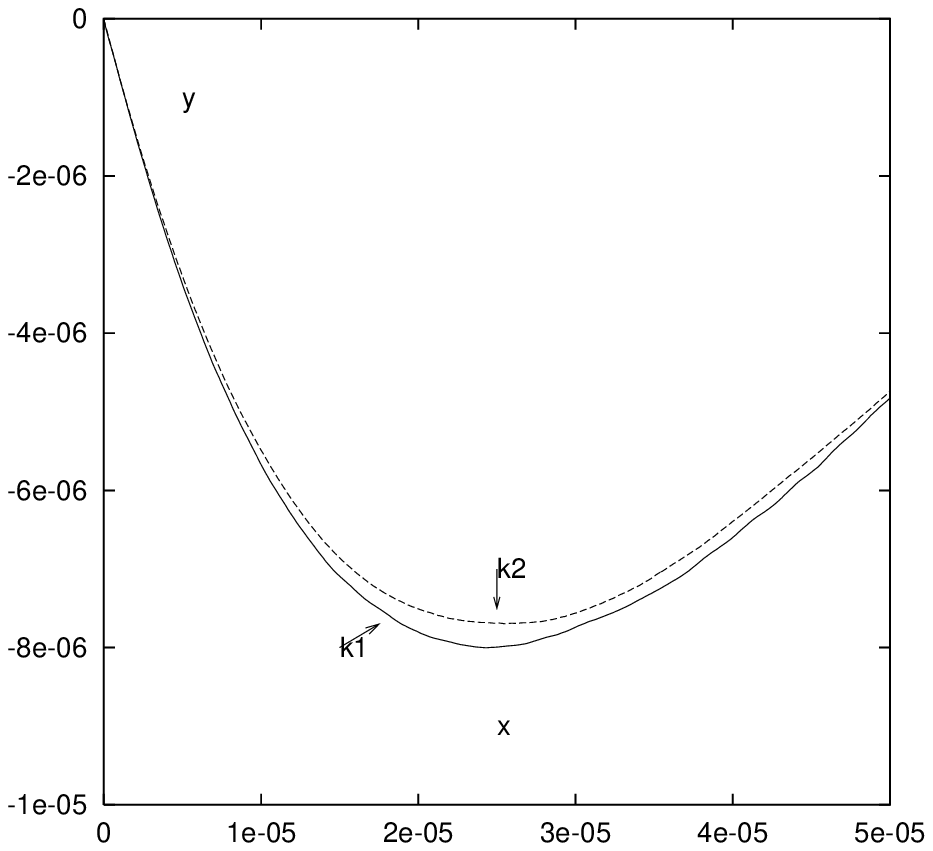}
 \caption{Stress ratio $q/p$ (left) and rescaled volumetric strain $\kappa\epsilon_v$ (right) vs.
rescaled axial strain $\kappa\epsilon_1$ for $\kappa=10^5$ and $\kappa=10^4$.
 \label{fig:varkn}}
\end{figure}

As the system, in regime I, is equivalent to a network of springs and plastic sliders (Fig.~\ref{fig:rheonet}),
type I strains are all inversely proportional to stiffness level $\kappa$, provided the compression that decreases $\kappa$ does not significantly affect the sample geometry. The curves pertaining to different $\kappa$ 
values coincide if expressed with stress ratios and variables $\kappa\epsilon$, as shown in Fig~\ref{fig:varkn}.
\subsubsection{Approach to the macroscopic limit}
\begin{figure}[!htb]
\centering
\includegraphics[width=6.5cm]{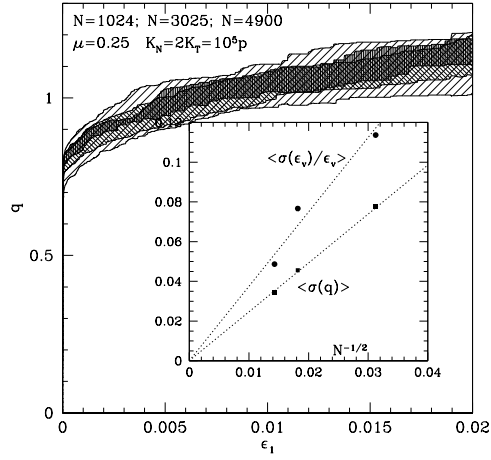}
\caption{\label{fig:courbstat}
Main plot: sample to sample average of $q$ versus $\epsilon_1$. Shaded regions extend to one
standard deviation about the average, with, in this order, darker and darker shades of gray for N=1024, 3025, 4900. Insert: regression of fluctuations, proportional to $N^{-1/2}$. Average standard deviations $\sigma(q)$ and $\sigma(\epsilon_v)$ over interval $0\le\epsilon_1\le 0.02$. 
}
\end{figure}

The staircase-shaped loading curves in regime II should approach in the large sample limit a smooth stress-strain curve, as
observed in very slow laboratory tests. To check for the approach of such a macroscopic limit, the 
average $\ave{q(\epsilon_1)}$ and the standard deviation $\sigma(q)(\epsilon_1)$ are computed as functions of
axial strain for sets of samples of three different sizes,  and the region of the
$\epsilon_1$ -- $q$ plane corresponding to $\ave{q}(\epsilon_1)-\sigma(q)\le q \le \ave{q}(\epsilon_1)+\sigma(q)$ is shaded on Fig.~\ref{fig:courbstat}, 
darker zones corresponding to larger N. Fluctuations about the average
curve decrease as the system size increases, and the insert shows that the standard deviation, as averaged over
the interval $0\le\epsilon_1\le 0.02$, regresses as $N^{-1/2}$. Thus  staircase curves get smoothed in the large system limit,
which implies that the ``stairs'' become increasingly small and numerous: as $N$ increases rearrangement events (microscopic
instabilities) become smaller and smaller, but more and more frequent. 
A similar regression is observed for the volumetric strain curve. 

Unlike the small type I response intervals observed within regime II, the stability range $q\le q_1$ of the 
initial, isotropic structure does not dwindle as the system size increases. As shown in Table~\ref{tab:q1}, 
\begin{table}
 \begin{tabular}{|c||c|c|c|}  \cline{1-4}
   \emph{SEM} & \bf{N=1024}       & \bf{N=3025} & \bf{N=4900} \\
   \hline
   $\langle q_1 \rangle$ & $0.750 \pm 0.050$ & $0.774 \pm 0.033$ &  $0.786 \pm 0.024$   \\
   \hline
 \end{tabular} 
\caption{Average and standard deviation of $q_1$ as obtained over 26 samples with N=1024, 10 samples with N=3025 and 6 samples with N=4900.}
 \label{tab:q1}
\end{table}
the initial regime I deviator interval even increases a little, approaching a finite limit as $N\to\infty$.

Our implementation of SEM involves no creation of new contacts (although this could be taken into account in a more
refined version). This approximation becomes exact in the rigid limit of $\kappa\to\infty$, because a finite
strain increment is necessary for additional contacts to close, while type I strains scale as $\kappa^{-1}$. 
The near coincidence of SEM and DEM approaches, the latter involving contact creations, shows that new contacts are indeed negligible for $\kappa=10^5$. 
$q_1$ thus represents the maximum deviator stress supported by 
the initial contact network, beyond which~\cite{McHe06}, due to contact sliding and opening, an instability or a 
``floppy mode'' appears. The hallmark of such instabilities is the negativity of the second-order work~\cite{McHe06,RC09}, \emph{viz.}
$$
\Delta^2W(\Delta\dep) = \Delta\dep\cdot\sti\cdot\Delta\dep 
$$
for some direction of displacement increment vector $\Delta\dep$. Vector $\Delta\dep$ comprises all increments of grain displacements 
and rotations, and $\sti$ is the stiffness matrix, which depends, \emph{via} the status of contacts, on the direction of $\Delta\dep$. 
$\Delta^2W(\Delta\dep)<0$ implies that the increment of contact forces resulting
from a small perturbation $\Delta\dep$ will accelerate the resulting motion, whence a spontaneous increase
of kinetic energy.
\subsubsection{Transition stress $q_1$ and the ``critical yield analysis'' approach}
\begin{figure}[!htb] 
\centering
\includegraphics[angle=270,width=8.5cm]{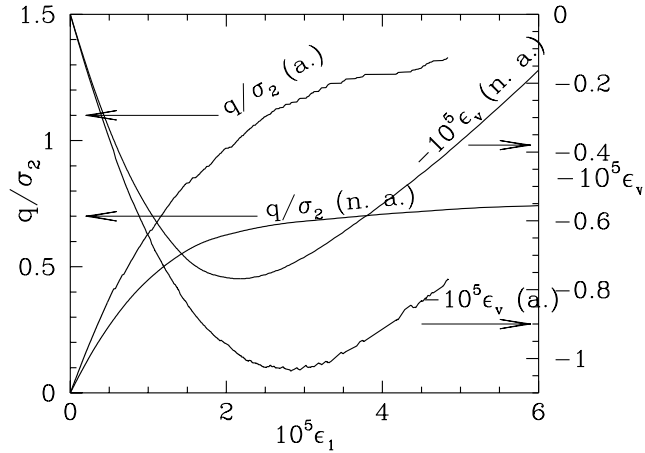}
\caption{Comparison of SEM calculation with the normal (parallel, curves marked ``n. a.'' for non associated) and 
the ``associated'' (dilatant, curves marked ``a.'') sliding rule in contacts in sample with 1024 disks.
\label{fig:compasso}}
\end{figure}

One may wonder whether $q_1$  marks the upper bound $q_u$ of the deviator interval for which contact forces balancing
the external load (i. e., statically admissible) and satisfying Coulomb's inequality (i. e., plastically admissible) can be found in the network. 
This is the ``critical yield analysis'' approach to failure in structural mechanics. It 
is known that $q_1$ and $q_u$ would coincide if the sliding in contacts where friction is fully mobilized 
implied dilatancy,  with an angle equal to the friction angle 
(the discrete analog of an ``associated'' flow rule). Fig.~\ref{fig:compasso} shows that $q_1$ is well below $q_u$.
With a dilatant sliding rule, the material response in biaxial compression would be stiffer, and initially (rather
paradoxically) more contractant, and the deviator would reach $1.3P$ (instead of about $0.8P$) before failure of the
initial contact network.
\subsubsection{Evolution of microscopic state variables}
\begin{figure}[htp!]
 \centering
 \psfrag{x}[c]{$\epsilon_1$}
 \psfrag{y1}[c]{$h$}
 \psfrag{y2}[r]{$\chi_s/N_c$}
 \psfrag{2e-05}[l]{\begin{small}$2.10^{-5}$\end{small}}
 \psfrag{4e-05}[l]{\begin{small}$4.10^{-5}$\end{small}}
 \psfrag{6e-05}[l]{\begin{small}$6.10^{-5}$\end{small}}
 \psfrag{8e-05}[l]{\begin{small}$8.10^{-5}$\end{small}}
 \psfrag{0.0001}[l]{\begin{small}$10^{-4}$\end{small}}
  \psfrag{0}[l]{\begin{small}$0$\end{small}}
\includegraphics[width=0.475\textwidth]{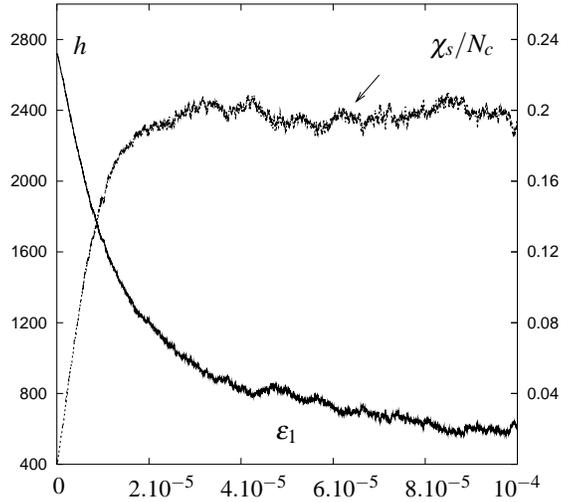}
 \caption{In a sample with N=3025, degree of force indeterminacy $h$ (solid line) and 
proportion of sliding intergranular contacts $\chi_s/N_c$ (dotted line), versus axial strain $\epsilon_1$.
\label{fig:micro}}
\end{figure}
As mentioned above, contact creation is negligible in regime I and the fabric evolution is essentially due to
contacts opening, mostly in the direction of extension (direction 2). As the initial coordination number is maximal, because of the absence of friction in the assembling process, very few contacts are gained in the direction of compression. In the initial state, all contacts only bear normal force components. Friction
mobilization is gradual, but the proportion of sliding contacts, as shown in Fig.~\ref{fig:micro} steadily increases from zero in regime I, and reaches an apparent plateau in regime II. This means that the interval of
elastic response is, strictly speaking, reduced to naught, even though the stress-strain curve can \emph{approximately} be described as elastic in a very small range. 
The appearance of sliding contacts can effectively reduce the degee of static indeterminacy in the system.
If the status is assumed to be fixed for all contacts, 
the Coulomb condition, satisfied as an equality, reduces the number of independent contact 
components from $2dN_c$ (in 2D) to $2N_c-\chi_s$, with $\chi_s$ the number of sliding contacts among a total of
$N_c$. It has been speculated~\cite{LJ00,Rothenburg09} that failing contact networks (regime II) should correspond
to vanishing force indeterminacy. The data of Fig.~\ref{fig:micro} provide evidence against such a prediction, as $h$ stabilizes to about 600, a moderate (10\% of the total number of degrees of freedom in a sample of 3025 disks), yet finite value. 
\subsection{3D triaxial tests}
The simulations reported here compare dense states A (high coordination number) and C (low coordination number).
State A is similar to the dense disk sample studied in the previous section, as both were initially assembled 
with frictionless grains. (A samples, once
packed under low pressure, are nevertheless compressed to the desired confining pressure with the value 
$\mu=0.3$ of the friction coefficient used in the triaxial tests~\cite{iviso1}). Pressure values correspond to glass beads,
and vary between $10$~kPa ($\kappa = 39000$) and 1~MPa ($\kappa = 1800$). We first check for the approach of
the quasistatic and the macroscopic limit in 3D, strain-rate controlled DEM simulations, then discuss the influence
of coordination number, and regimes I and II, in the light of the previous 2D study.

\subsubsection{Reproducibility, quasistatic limit}
\begin{figure}[!htb]
\begin{minipage}[l]{.45\textwidth}
\includegraphics[angle=270,width=6cm]{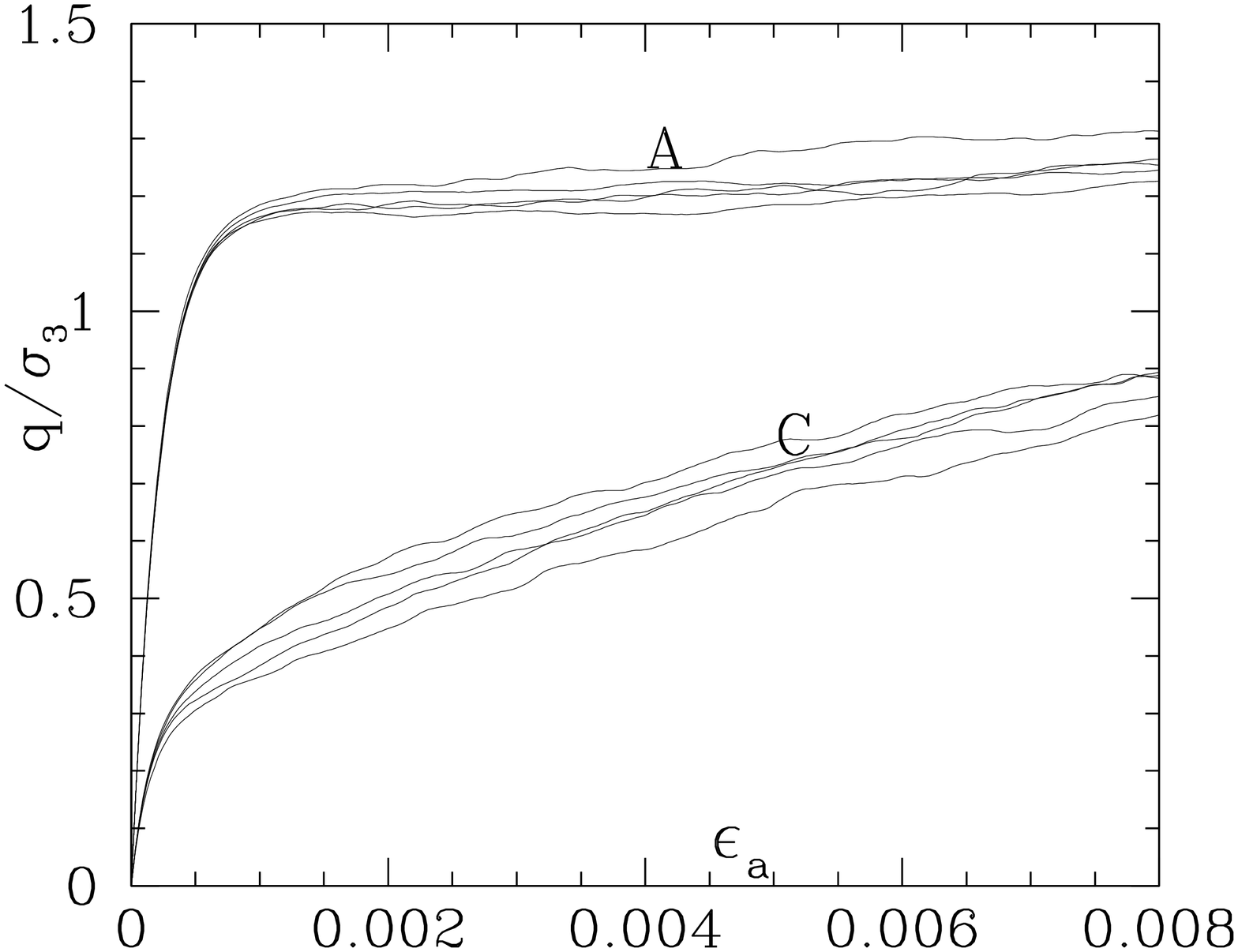}
\end{minipage}
\hfil
\begin{minipage}[l]{.45\textwidth}
\includegraphics[width=6cm]{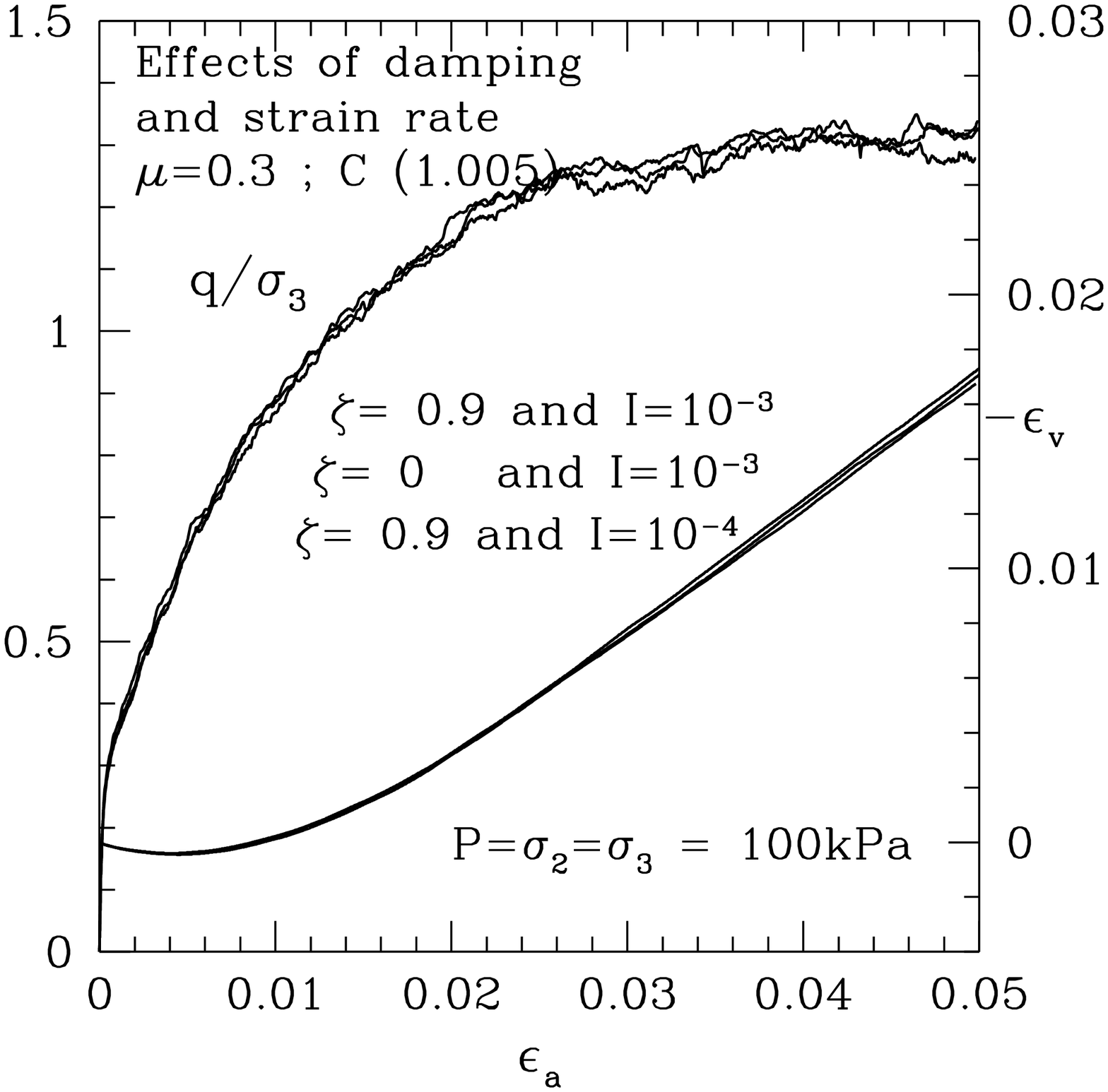}
\end{minipage}
\caption{Left: small strain part of $q(\epsilon_a)$ curves for 5
different samples of each type, A (top curves) and C (bottom ones) with $N=4000$ beads. Right:
$q(\epsilon_a)$ and $\epsilon_v(\epsilon_a)$ curves in one type C sample for the different values of
$\zeta$ and $I$ indicated.
\label{fig:triaxcomp2}
}
\end{figure}

As the system size increases, 
sample to sample fluctuations should regress, as checked in 2D (Fig.~\ref{fig:courbstat}).
Our 3D results are based on 5 samples of 4000 beads of each type, and Fig.~\ref{fig:triaxcomp2} checks for 
stress-strain curve reproducibility in both A and C cases, for small axial strains. Thanks to
the fully periodic boundary conditions~\cite{iviso1}, the macroscopic mechanical behavior is quite
well defined with $N=4000$. The approach to the quasistatic limit, in SRC tests can be assessed on checking
for the innocuousness of the dynamical parameters, i.e., inertial number $I$, 
and reduced damping parameter $\zeta$. $\zeta$ is defined as the
ratio of the viscous damping constant in a contact to its critical level, given the instantaneous value of the
stiffness constant. We found it convenient to use a constant value of $\zeta$ in our simulations, as in~\cite{iviso1}. Fig.~\ref{fig:triaxcomp2} also shows that provided inertial number $I$, characterizing
dynamical effects, is small enough, both $I$ and $\zeta$ become irrelevant.
Fig.~\ref{fig:triaxcomp2} shows that the quasistatic limit is correctly
approached for $I\le 10^{-3}$, quite a satisfactory result, given that usual laboratory tests with $\dot \epsilon_a \sim 10^{-5}$~s$^{-1}$ correspond to $I\le 10^{-8}$.
\subsubsection{Influence of initial coordination number}
\begin{figure}[!htb]
\centering
\includegraphics[width=6cm]{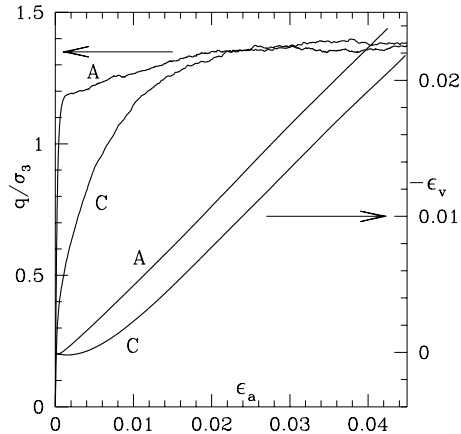}
\caption{ 
$q(\epsilon_a)$ (left scale)  and $\epsilon_v(\epsilon_a)$ (right scale) curves for A and C 
states under $P=100$~kPa. Averages over 5 samples of 4000 spherical grains.
\label{fig:triaxcomp}}
\end{figure}

Fig.~\ref{fig:triaxcomp} compares
the behavior of initial states A and C, in triaxial compression
with $P=100$~kPa ($\kappa \simeq 6000$).
Although, conforming to the traditional view that the peak deviator stress is determined by the initial sample density, maximum $q$ values are very nearly identical in systems A and C, the mobilization of internal friction is
much more gradual for C. For A, the initial rise of deviator $q$ for small axial strain is quite steep, and the volumetric
strain variation  becomes dilatant almost immediately, for $\epsilon_a \sim 10^{-3}$.
In~\cite{iviso3} it was shown that measurements of elastic moduli provide information on coordination numbers.
It is thus conceivable to infer the rate of deviator increase as a function of axial strain from very small
strain ($\sim 10^{-5}$ or below~\cite{Tat104,iviso3}) elasticity.
Most experimental curves obtained on sands, which do not exhibit $q$ maxima or dilatancy before $\epsilon_a \sim 0.01$, are closer to C ones. 
However, some measurements on glass bead samples~\cite{Emam06} do show fast rises of $q$ at small strains, somewhat intermediate between numerical results of types A and C.
\subsubsection{Influence of contact stiffness}
\begin{figure}[!htb]
\centering
\includegraphics[width=6cm]{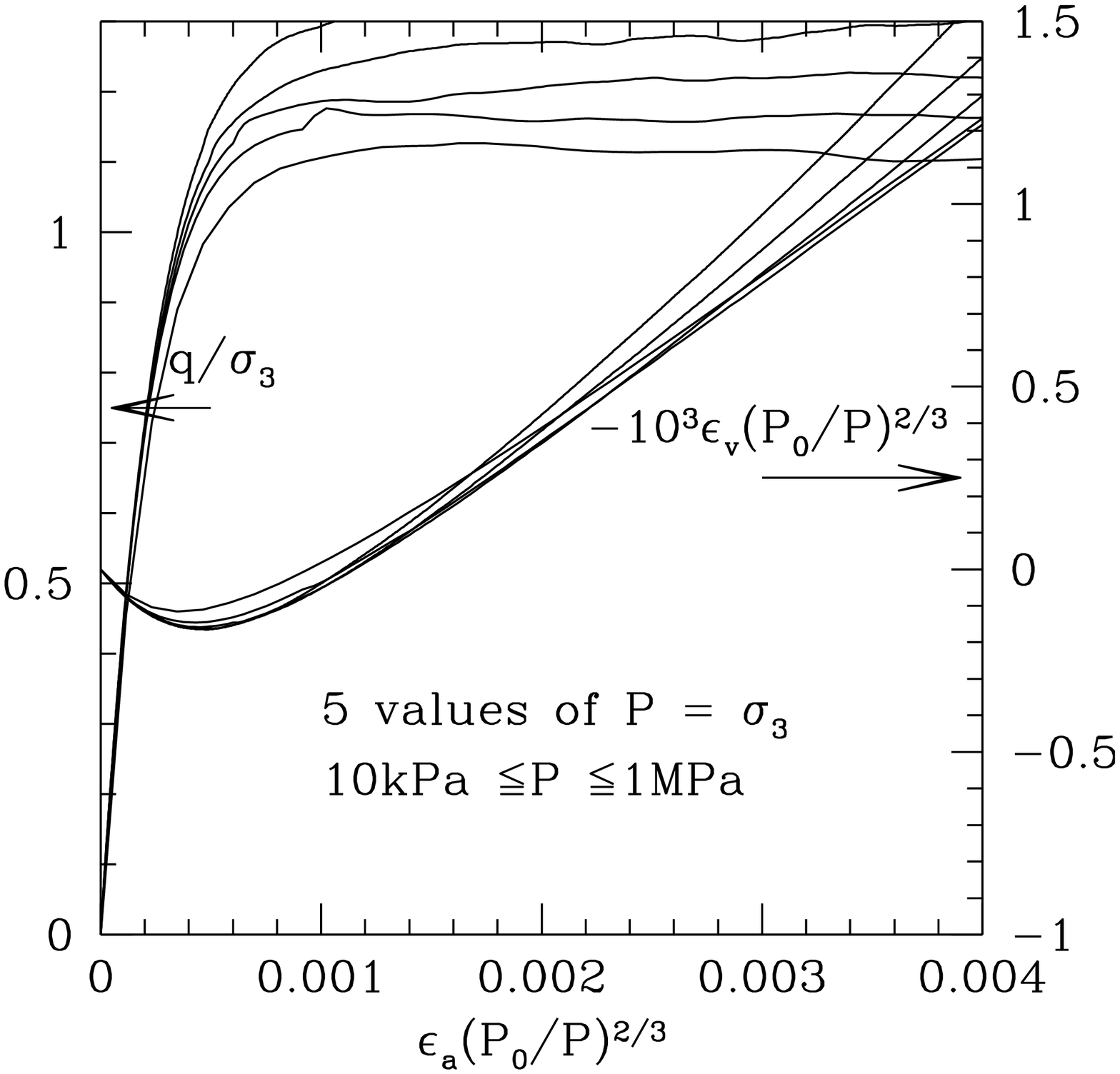}
\hfil
\includegraphics[width=6cm]{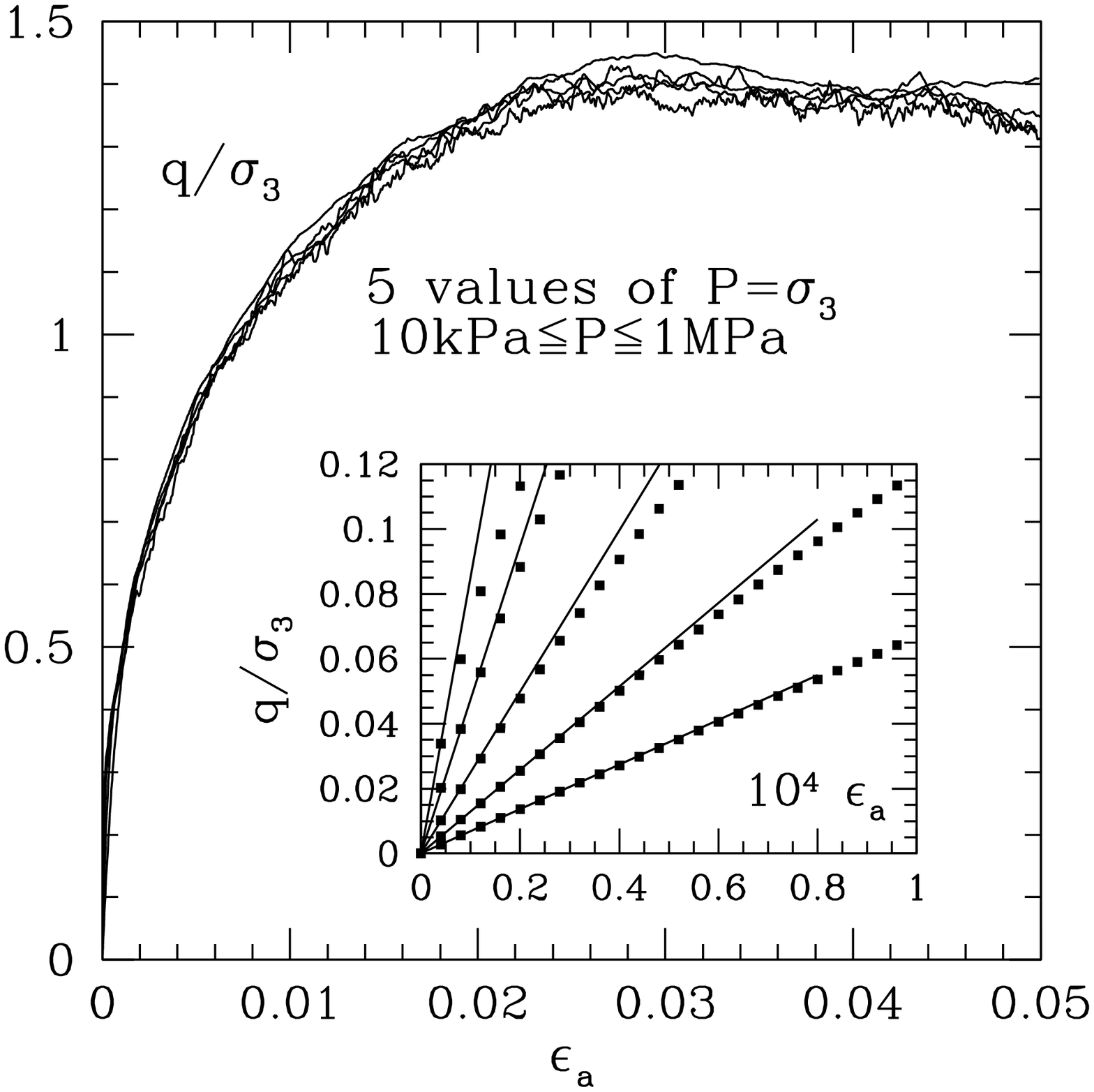}
\caption{Left: $q(\epsilon_a)/P$ and $\epsilon_v(\epsilon_a)$ curves for one A sample and different
$P$ values. Strains on scale $(P/P_0)^{2/3}\propto \kappa^{-1}$, $P_0=100$~kPa. Right:
$q(\epsilon_a)/P$ for the same $P$ values in one C sample. 
Inset: detail with blown-up $\epsilon$ scale, straight lines corresponding to Young moduli in isotropic state.
\label{fig:dessPP2}}
\end{figure}

The small strain (say $\epsilon_a\le 5.10^{-4}$) interval for A samples, with its fast $q$ increase, is in regime I, as one might expect from 2D results on disks. 
This is readily checked on changing the confining pressure. Fig.~\ref{fig:dessPP2} shows 
the curves for triaxial compressions at different $P$ values
(separated by a factor $\sqrt{10}$) from $10$~kPa to $1$~MPa,
with a rescaling of the strains by the stiffness parameter $\kappa$,
in one A sample. 
Their coincidence for $q/P\le 1$ evidences a wide deviator range in regime I. 
For larger strains, curves separate on this scale, and tend to collapse together if $q/P$, $\epsilon_v$ are simply plotted versus
$\epsilon_a$. The strain
dependence on stress ratio is independent from contact stiffness. This different sensitivity to pressure
is characteristic of regime II.
Fig~\ref{fig:dessPP2} also shows that it applies to C samples almost throughout the investigated
range, down to small deviators (a behavior closer to usual
experimental results than type A configurations). At the origin (close to the
initial isotropic state, see inset on fig.~\ref{fig:dessPP2}, right plot), the
tangent to the curve is given by the elastic (Young) modulus of the
granular material, $E_m$, and therefore $q/P$ scales with $\kappa$, but curves 
quickly depart from this behaviour (around $q=0.2P$). The approximately elastic range~\cite{iviso3} is quite
small, as observed in experiments~\cite{Tat104,dBGSC99,KJ02}.
\subsubsection{Calculations with a fixed contact list}
\begin{figure}[!htb]
\centering
\includegraphics[height=7cm,angle=270]{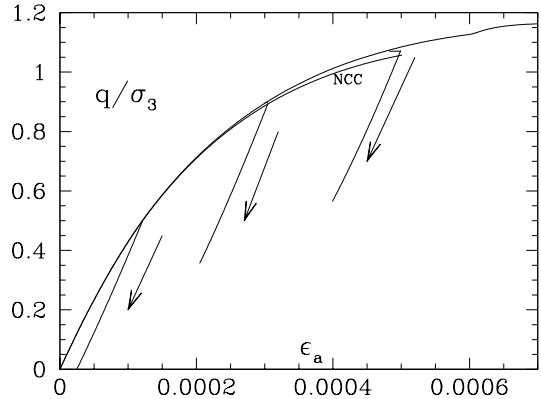}
\caption{Very small strain part of $q(\epsilon_a)$ curve in one A sample, showing beginning of unloading curves (arrows).
Curve marked  NCC was obtained on calculating the evolution of the same sample without any contact creation. \label{fig:revers}
}
\end{figure}
Within regime I, the mechanical properties of the material can be
successfully
predicted on studying the response of one given set of contacts.
Those might slide or open, but the very few new contacts that are
created can be neglected.
To check this in simulations, 
one may restrict at each time step the search for interacting grains
to the list of initially contacting pairs.
Fig.~\ref{fig:revers} compares such a procedure to the complete
calculation. 
The curve marked ``NCC'' for \emph{no
  contact creation} is indistinguishable from the other one for $q\ge 0.8$. We thus check that, in regime I,
the macroscopic behavior is essentially determined by the response of a fixed contact network.
\subsubsection{Type I strains and elastic reponse}
Fig.~\ref{fig:revers} also shows that the small strain response of A samples, within regime I, close to the initial state, is already irreversible: type I strains are not elastic. 
An approximately elastic behavior is only
observed for very small strains, as depicted in the inset of Fig.~\ref{fig:dessPP2} (right part). In this small
interval near the initial equilibrium configuration, the
stress-strain curve is close to its initial tangent, defined by the elastic modulus. Moduli~\cite{iviso3} can
be calculated from the stiffness matrix of contact networks. One may also check that the unloading curves
shown on Fig.~\ref{fig:revers} (and the ones of Fig.~\ref{fig:dessmarfro} in 2D as well) comprise a small,
approximately elastic part, with the relevant elastic modulus (the Young modulus for a triaxial test at constant
lateral stress) defining the initial slope. At the microscopic level, a small elastic response is retrieved 
upon reversing the loading direction because contacts stop sliding. The elastic range is strictly included in the
larger range of type I behavior.
\subsubsection{Fluctuations and length scale}
Finally, let us note that regimes I and II also differ by the importance of sample to sample fluctuations: curves in Fig.~\ref{fig:triaxcomp2} (left plot) pertaining to the different samples of type A or C are confused as long as $q\le 1.1 P$ (case A) or $q\le 0.3P$ (case C), which roughly corresponds to the transition from regime I to regime II. Larger fluctuations imply that the characteristic length scale associated with the displacement field
(correlation length) is larger in regime II. Whether and in what sense rearrangements triggered by instabilities
in regime II, in a material close to the rigid limit (large $\kappa$), can be regarded as local events is still 
an open issue.
\section{CONCLUSION}
Numerical studies thus reveal that the two regimes, in which the origins of
strain differ, exhibit contrasting properties. 
Although the reported studies in 2D and 3D differ in many respects (linear versus Hertzian contacts,
wall versus periodic boundaries, SIC versus SRC DEM), the same phenomena were observed in both cases.
Regime I corresponds to the stability range of a given contact structure. It is larger in
highly coordinated systems. It is observed in the beginning of monotonic loading tests, in which the deviator
stress increases from an initial isotropic configuration, and also after changes in the direction
of load increments (hence a loss in friction mobilization). Strains, for a given stress level, 
are then inversely proportional to contact stiffnesses. The deviator range in regime I, $q\le q_1$, in 
usual monotonic tests, is stricly larger than the small elastic range, but strictly smaller than the 
maximum deviator. It does not vanish in the limit of large systems, unlike in the singular case of
rigid, frictionless particle assemblies~\cite{CR2000,PR08b}.
Regime I is limited by the occurrence of elastoplastic instabilities in the contact network and 
does not coincide with the prediction of the critical yield approach. In regime I, the work of the externally 
applied load is constantly balanced by the one of contact forces, so that the kinetic energy approaches zero in
the limit of slow loading rates. A remarkable consequence is that the instability condition based on
the negativity of the macroscopic second-order work~\cite{PLD09} is never fulfilled, as macroscopic and microscopic works coincide, and the latter is positive. In regime II, network rearrangements are triggered by instabilities and some bursts of kinetic energy are observed~\cite{staron02b}.
Larger fluctuations witness longer-ranged correlations in the displacements. The microscopic origin of 
macroscopic strains, which are independent on contact elasticity for usual stiffness levels $\kappa$, lies in the
geometry of grain packings. 

On attempting to predict a macroscopic mechanical response from packing geometry and contact
laws, the information about which kind of strain should dominate is crucial. 

A promising perspective is the study of correlated motions associated with rearrangement events. 
\bibliographystyle{aipproc}  

\bibliography{granu3}

\end{document}